\documentclass[prb,twocolumn]{revtex4}
\usepackage{amsfonts}
\usepackage{amsmath}
\usepackage{amssymb}
\usepackage{epsfig}
\usepackage{graphicx}

\begin{document}

\title{Bose-Einstein condensation of magnons in Cs$_{2}$CuCl$_{4}$: a dilute
gas limit near the saturation magnetic field}
\author{D.L. Kovrizhin$^{1,3}$, V. Yushankhai$^{2,4}$, L. Siurakshina$^4$}
\affiliation{$^1$Max-Planck-Institute f\"{u}r Physik Komplexer Systeme, N\"{o}thnitzer
Str. 38, 01187 Dresden, Germany\\
$^2$Max-Planck-Institute f\"{u}r Chemical Physik Fester Stoffe, N\"{o}%
thnitzer Str. 40, 01187 Dresden, Germany \\
$^3$Russian Research Centre Kurchatov Institute, Kurchatov sq. 1, 123182
Moscow, Russia\\
$^4$Joint Institute for Nuclear Research, 141980 Dubna, \ Russia}

\begin{abstract}
Based on a realistic spin Hamiltonian for a frustrated quasi-two dimensional
spin-$1/2$ antiferromagnet Cs$_{2}$CuCl$_{4}$, a three-dimensional spin
ordering in the applied magnetic field $B$ near the saturation value $B_{c}$
is studied within the magnon Bose-Einstein condensation (BEC) scenario. With
the use of a hard-core boson formulation of the spin model, a strongly
anysotropic magnon dispersion in Cs$_{2}$CuCl$_{4}$ is calculated. In the
dilute magnon limit near $B_{c}$, the hard-core boson constraint is resulted
in an effective magnon interaction which is treated in the Hartree-Fock
approximation. The critical temperature $T_{c}$ is calculated as a function
of a magnetic field $B$ and compared with the phase boundary $T_{c}\left(
B\right) $ experimentally determined in Cs$_{2}$CuCl$_{4}$ [Phys. Rev. Lett.
\textbf{95}, 127202 (2005)].
\end{abstract}

\maketitle

\section{Introduction}

The insulator Cs$_{2}$CuCl$_{4}$ is a quasi-two dimensional spin-$1/2$
antiferromagnet (AFM) composed of triangular lattices (\textit{bc}-plane)
which are weakly coupled along crystallographic \textit{a}-direction. \cite%
{Coldea03} In spite of a low spin, low dimensionality and geometrical
frustration of spin interactions, a three-dimensional (3D) magnetic
long-range ordering occurs in Cs$_{2}$CuCl$_{4}$ at sufficiently low
temperatures $T\lesssim T_{N}\simeq 0.6$K. In the ordered state, magnetic
moments lie in an easy \textit{bc-}plane and form an incommensurate spiral
structure.\cite{Coldea03} An easy-plane spin anisotropy in Cs$_{2}$CuCl$_{4}$
is produced by a Dzyaloshinskii-Moriya (DM) interaction with a DM vector $%
\mathcal{\vec{D}}\parallel \vec{a}$. The anisotropy breaks SU$\left(
2\right) $ symmetry, however, U$\left( 1\right) $ symmetry corresponding to
spin rotations around $\mathcal{\vec{D}}$ vector is still present. The
applied magnetic field $\vec{B}\parallel \vec{a}$ preserves U$\left(
1\right) $ symmetry and causes the spiral spin order to develop into a cone
spin structure with increasing $B$: a longitudinal spin component grows with 
$B$, while a spiral transverse component decreases steadily and disappears
completely at a saturation field $B_{c}$. For higher fields, $B>B_{c}$, the
Zeeman interaction overcomes the antiferromagnetic spin coupling and the
system enters a field-induced ferromagnetic state.\cite{Coldea02} In this
state, \textit{ferromagnetic} magnons become elementary spin exciatations
with a gapped spectrum and a quadratic energy dispersion near magnon band
minima. In the opposite direction, i.e. by starting from a high field $%
B>B_{c}$ and passing through $B_{c}$, one finds a reversed behavior: the
field-induced magnon gap dissapears at $B_{c}\,$\ and a spiral transverse
magnetic ordering develops below $B_{c}$. This ordered state is
characterized by a gapless spin excitation spectrum with a linear low-energy
dispersion.

Such a phase transition observed\cite{Coldea02,Radu05} in Cs$_{2}$CuCl$_{4}$
in the applied magnetic field $\vec{B}\parallel \vec{a}$ in a vicinity of $%
B_{c}$ can be regarded as a Bose-Einstein condensation (BEC) of a dilute gas
of magnons.\cite{Matsubara56,Batyev84} This suggestion was confirmed
experimentally and supported theoretically in Ref. \onlinecite{Radu05}. In
this context, two particular features of Cs$_{2}$CuCl$_{4}$ have to be
mentioned. First, because of a rather weak dominant exchange coupling $J$ $%
\simeq 4.3$K in this material, an exceptionally low saturation field $%
B_{c}\simeq 8.5$T can be easily achieved. Second, because of the extremely
strong anisotropy of the ferromagnetic magnon spectrum in Cs$_{2}$CuCl$_{4}$%
, 3D quadratic magnon dispersion is seen at very low energies $E<E^{\ast
}\simeq 50$mK, and for $E>E^{\ast }$ the actual magnon spectrum is of 2D
character. Therefore, an access to very low temperatures $T\sim 50$mK is
required to be as close as possible to the asymptotic regime where universal
3D scaling laws are expected to hold.

Cs$_{2}$CuCl$_{4}$ is an easy-plane quantum AFM with a gapless spin
excitation spectrum at zero magnetic field.\cite{Coldea03} Up to now, most
of studies of a magnon (triplon) BEC have been reported\cite%
{Oosawa99,Nikuni00,Misguich04,Jaime04,Sebastian05} on TlCuCl$_3$ and BaCuSi$%
_2$O$_6$ that belong to a more rare class of gapped AFMs. In these
materials, a transverse spin long-range ordering occurs in the field range $%
B_{c1}< B < B_{c2}$. In TlCuCl$_3$ and BaCuSi$_2$O$_6$, a saturation field $%
B_{c2}$ is rather high $\sim$50T, which makes an observation of a magnon BEC
at $B_{c2}$ a complicated matter, while lower fields near $B_{c1}$ are
accessible. Within the 3D magnon BEC scenario,\cite%
{Giamarchi99,Nohadani04,Kawashima04} near $B_{c1}$ the ordering temperature $%
T_{c }\left( B\right)$ is described by a power-law $T_{c}\left( B\right)
\sim\left( B - B_{c}\right) ^{1/\phi}$ with a universal critical exponent $%
\phi_{BEC}=3/2$. A few experimental findings show, however, that the
observed in TlCuCl$_3$ phase transition deviates from a pure magnon BEC:
first, an anisotropic spin coupling (of unknown nature) might break the
axial U$\left( 1\right) $ symmetry of the system and produce a small but
finite spin gap\cite{Glazkov04,Kolezhuk04,Sirker05} in the ordered state at $%
B > B_{c1}$ and, second, the reported critical exponent $\phi$ is somewhat
larger than predicted by theory.\cite{Nikuni00} Magnetic properties of BaCuSi%
$_2$O$_6$ are suggested to be well described by a quasi-two dimensional
isotropic Heisenberg model and the reported\cite{Sebastian05} recently
convergence of the measured exponent $\phi$ to the predicted value $%
\phi_{BEC}=3/2$ is in accord with the 3D magnon BEC scenario.

In this paper a theoretical description of the magnon BEC transition in Cs$%
_{2}$CuCl$_{4}$ near the saturation field is presented with more details as
compared to the previous work.\cite{Radu05} In section II, a model spin
Hamiltonian applicable to Cs$_{2}$CuCl$_{4}$ is discussed and rewritten in
the hard-core boson representation. Two branches of the bare magnon
dispersion and their densities of states are calculated. In section III, the
hard-core boson constraint is treated in ladder approximation and the
Bethe-Salpeter equation is solved to calculate the effective magnon
interaction in the dilute gas limit. In the next section IV, the obtained
effective interaction is considered in a mean-field approximation and the
leading effects of magnon interaction are extracted and discussed. In
section V, the critical temperature $T_{c}\left( B\right) $ as a function of
applied magnetic field is calculated and compared both with the predicted
universal behavior of the phase boundary\cite%
{Giamarchi99,Nohadani04,Kawashima04} near $B_{c}$ and with experimental
data. Short concluding remarks can be found in section VI.

\section{Spin Hamiltonian and magnon dispersion}

The spin Hamiltonian $\mathcal{H}$ of Cs$_{2}$CuCl$_{4}$ involves the
isotropic exchange $\mathcal{H}_{0}$, the DM anisotropic term $\mathcal{H}%
_{DM}$ and the Zeeman energy $\mathcal{H}_{B}$. By using the notations of
Ref. \onlinecite{Veillette05}, the isotropic part $\mathcal{H}_{0}$ can be
written as%
\begin{multline}
\mathcal{H}_{0}=\sum_{\mathbf{R}}[J\vec{S}_{\mathbf{R}}\vec{S}_{\mathbf{R}%
+\tau _{1}+\tau _{2}}+J^{\prime }(\vec{S}_{\mathbf{R}}\vec{S}_{\mathbf{R}%
+\tau _{1}}+\vec{S}_{\mathbf{R}}\vec{S}_{\mathbf{R}+\tau _{2}}) \\
+J^{\prime \prime }\vec{S}_{\mathbf{R}}\vec{S}_{\mathbf{R}+\tau _{3}}],
\label{eq1}
\end{multline}%
where the nearest neighbor vectors $\tau _{1}$ and $\tau _{2}$ are indicated
in Fig. 1 and the out-of-plane vector $\tau _{3}$ connects spins on vertical
bonds between adjacent layers. A small relative shift of adjacent layers is
neglected. 
\begin{figure}[tbh]
\epsfig{file=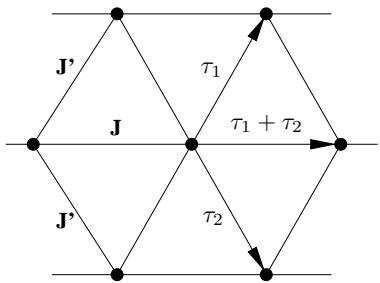,width=5cm} .
\caption{Fragment of the triangular lattice (\textit{bc}-plane) with
nearest-neighbor exchange couplings, $J$ and $J^{\prime }\approx J/3$. The
Dzyaloshinskii-Moriya coupling $\mathcal{D}$ is on the same zig-zag bonds as
the exchange $J^{\prime }$}
\label{fig:1}
\end{figure}
Within each layer, the dominant exchange $J$ distinguishes a chain direction
and the DM interaction is situated on interchain zig-zag bonds%
\begin{equation}
\mathcal{H}_{DM}=\sum_{\mathbf{R}}\left( -1\right) ^{n}\mathcal{\vec{D}}\vec{%
S}_{\mathbf{R}}\times \lbrack \vec{S}_{\mathbf{R}+\tau _{1}}+\vec{S}_{%
\mathbf{R}+\tau _{2}}].  \label{eq2}
\end{equation}%
The crystallographic $\vec{a}$-direction is chosen to be $\vec{z}$ \ axis
and because $\mathcal{\vec{D}}=\left( 0,0,\mathcal{D}\right) $ the \textit{bc%
}-plane becomes the easy plane for the ground state spin alignment in zero
magnetic field. In Eq. (\ref{eq2}) a layer index $n$ is introduced to show
that DM vectors alternate between adjacent layers. In the present study we
consider the applied magnetic field $\vec{B}\parallel \vec{a}$ and Zeeman
energy reads%
\begin{equation}
\mathcal{H}_{B}=-g\mu _{B}B\sum_{\mathbf{R}}{S}_{\mathbf{R}}^{z},
\label{eq3}
\end{equation}%
where $g\simeq 2.2$ and $\mu _{B}$ is the Bohr magneton. The spin coupling
parameters determined with high accuracy by neutron scattering measurements%
\cite{Coldea02} are: $J=4.34$K, $J^{\prime }/J=0.34$, $J^{\prime \prime
}/J=0.045$ and $\mathcal{D}/J=0.053$. The total Hamiltonian $\mathcal{H}$
has U$\left( 1\right) $ symmetry arising from spin rotations around $\vec{z}$
axis. In Cs$_{2}$CuCl$_{4}$ the saturation field $B_{c}\simeq 8.5$T and for $%
B>B_{c}$ at zero temperature the spins are fully polarized. To consider the
magnetic phase transition in the vicinity of $B_{c}$ and at low $T$, we use
the hard-core boson representation for spin-$1/2$ operators. Two types of
bosons $a_{\mathbf{R}}$ and $b_{\mathbf{R}^{\prime }}$ are introduced for
even and odd layers, respectively. This is given with a replacement ${S}_{%
\mathbf{R}}^{+}$ $\rightarrow $ $a_{\mathbf{R}}$, ${S}_{\mathbf{R}}^{-}$ $%
\rightarrow $ $a_{\mathbf{R}}^{+}$ and ${S}_{\mathbf{R}}^{z}=1/2-a_{\mathbf{R%
}}^{+}a_{\mathbf{R}}$ in even layers and the same transformation leads to $%
b_{\mathbf{R}^{\prime }}$ for $\mathbf{R}^{\prime }$ belonging to odd
layers. Such a replacement must be complemented by a constraint of no sites
with boson occupancies higher than unity. The constraint is satisfied by
adding to $\mathcal{H}$ an infinite on-site repulsion, ${\mathcal{U}}%
\rightarrow \infty $, among the bosons%
\begin{equation}
\mathcal{H}_{\mathcal{U}}^{\left( a\right) }+\mathcal{H}_{\mathcal{U}%
}^{\left( b\right) }=\mathcal{U}\sum_{\mathbf{R}}a_{\mathbf{R}}^{+}a_{%
\mathbf{R}}^{+}a_{\mathbf{R}}a_{\mathbf{R}}+\mathcal{U}\sum_{\mathbf{R}%
^{\prime }}b_{\mathbf{R}^{\prime }}^{+}b_{\mathbf{R}^{\prime }}^{+}b_{%
\mathbf{R}^{\prime }}b_{\mathbf{R}^{\prime }}.  \label{eq4}
\end{equation}%
In the reciprocal space with an ortogonal basis we choose the Brillouin zone
with wave-vectors restricted to $0\leqslant q_{x}<2\pi $, $0\leqslant
q_{y}<4\pi $, and $0\leqslant q_{z}<2\pi $, and define Fourier transforms of
boson operators%
\begin{equation}
a_{\mathbf{q}}=\frac{1}{\sqrt{\mathcal{N}}}\sum_{\mathbf{R}}a_{\mathbf{R}%
}e^{-i\mathbf{qR}},\ \ \ b_{\mathbf{q}}=\frac{1}{\sqrt{\mathcal{N}}}\sum_{%
\mathbf{R}^{\prime }}b_{\mathbf{R}^{\prime }}e^{-i\mathbf{qR}^{\prime }},
\label{eq5}
\end{equation}%
where $\mathcal{N}=N/2$ and $N$ is the total number of the spin-$1/2$
lattice sites in the system. Fourier transforms of spin interactions now read%
\begin{equation}
\begin{array}{c}
J_{\mathbf{q}}=J\cos q_{x}+2J^{\prime }\cos \left( q_{x}/2\right) \cos
\left( q_{y}/2\right) , \\ 
\\ 
J_{\mathbf{q}}^{\prime \prime }=J^{\prime \prime }\cos \left( q_{z}/2\right)
, \\ 
\\ 
\mathcal{D}_{\mathbf{q}}=2\mathcal{D}\sin \left( q_{x}/2\right) \cos \left(
q_{y}/2\right) .%
\end{array}
\label{eq6}
\end{equation}%
Two branches of 2D noninteracting magnons are described by the energy
dispersions%
\begin{equation}
\varepsilon _{\mathbf{q}}^{a,b}=J_{\mathbf{q}}\mp \mathcal{D}_{\mathbf{q}%
}+\Delta \varepsilon ,  \label{eq7}
\end{equation}%
where $\Delta \varepsilon =g\mu _{B}B-\left( J+2J^{\prime }+J^{\prime \prime
}\right) .$ With these notations, the bilinear part of the total Hamiltonian
takes the form%
\begin{multline}
\mathcal{H}_{bil}=\sum_{\mathbf{q}}[\varepsilon _{\mathbf{q}}^{a}a_{\mathbf{q%
}}^{+}a_{\mathbf{q}}+\varepsilon _{\mathbf{q}}^{b}b_{\mathbf{q}}^{+}b_{%
\mathbf{q}} \\
+J_{\mathbf{q}}^{\prime \prime }\left( a_{\mathbf{q}}^{+}b_{\mathbf{q}}+b_{%
\mathbf{q}}^{+}a_{\mathbf{q}}\right) ],  \label{eq8}
\end{multline}%
and leads to two 3D magnon branches, $\mathcal{A}$ and $\mathcal{B}$ with
bare dispersions%
\begin{equation}
\varepsilon _{\mathbf{q}}^{\mathcal{A},\mathcal{B}}=J_{\mathbf{q}}\mp 
\mathrm{sign}\mathcal{D}_{\mathbf{q}}\sqrt{\mathcal{D}_{\mathbf{q}}^{2}+J_{%
\mathbf{q}}^{\prime \prime 2}}+\Delta \varepsilon .  \label{eq9}
\end{equation}%
The degenerate minima $\varepsilon _{\mathbf{Q}_{1}}^{\mathcal{A}%
}=\varepsilon _{\mathbf{Q}_{2}}^{\mathcal{B}}$ are at $\mathbf{Q}_{1}=\left(
\pi +\delta _{1},0,0\right) $ for branch $\mathcal{A}$ and at $\mathbf{Q}%
_{2}=\left( \pi -\delta _{2},2\pi ,0\right) $ for branch $\mathcal{B}$.
Without loosing precision we can use $\delta _{1}\approx \delta _{2}\simeq
\delta =2\arcsin \left( J^{\prime }/2J\right) $. The bilinear part of $%
\mathcal{H}$ now reads%
\begin{equation}
\mathcal{H}_{bil}=\sum_{\mathbf{q}}[(E_{\mathbf{q}}^{\mathcal{A}}-\mu _{0})%
\mathcal{A}_{\mathbf{q}}^{+}\mathcal{A}_{\mathbf{q}}+(E_{\mathbf{q}}^{%
\mathcal{B}}-\mu _{0})\mathcal{B}_{\mathbf{q}}^{+}\mathcal{B}_{\mathbf{q}}],
\label{eq10}
\end{equation}%
where%
\begin{equation}
\left( 
\begin{array}{c}
\mathcal{A}_{\mathbf{q}} \\ 
\mathcal{B}_{\mathbf{q}}%
\end{array}%
\right) =\left( 
\begin{array}{cc}
\alpha _{\mathbf{q}} & \beta _{\mathbf{q}} \\ 
-\beta _{\mathbf{q}} & \alpha _{\mathbf{q}}%
\end{array}%
\right) \left( 
\begin{array}{c}
a_{\mathbf{q}} \\ 
b_{\mathbf{q}}%
\end{array}%
\right) ,  \label{eq11}
\end{equation}%
\begin{equation}
\begin{array}{c}
\alpha _{\mathbf{q}}=\sqrt{\dfrac{1}{2}\left[ 1+\dfrac{\left\vert \mathcal{D}%
_{\mathbf{q}}\right\vert }{\sqrt{\mathcal{D}_{\mathbf{q}}^{2}+J_{\mathbf{q}%
}^{\prime \prime 2}}}\right] }, \\ 
\\ 
\beta _{\mathbf{q}}=-\mathrm{sign}\mathcal{D}_{\mathbf{q}}\sqrt{\dfrac{1}{2}%
\left[ 1-\dfrac{\left\vert \mathcal{D}_{\mathbf{q}}\right\vert }{\sqrt{%
\mathcal{D}_{\mathbf{q}}^{2}+J_{\mathbf{q}}^{\prime \prime 2}}}\right] },%
\end{array}
\label{eq12}
\end{equation}%
and%
\begin{equation}
\begin{array}{c}
E_{\mathbf{q}}^{\mathcal{A}}=\varepsilon _{\mathbf{q}}^{\mathcal{A}%
}-\varepsilon _{\mathbf{Q}_{1}}^{\mathcal{A}}, \\ 
\\ 
\ E_{\mathbf{q}}^{\mathcal{B}}=\varepsilon _{\mathbf{q}}^{\mathcal{B}%
}-\varepsilon _{\mathbf{Q}_{2}}^{\mathcal{B}}.%
\end{array}
\label{eq13}
\end{equation}%
In Eq. (\ref{eq10}), the bare chemical potential of magnons $\mu
_{0}=-[\varepsilon _{\mathbf{Q}_{1}}^{\mathcal{A}}+\Delta \varepsilon
]=-[\varepsilon _{\mathbf{Q}_{2}}^{\mathcal{B}}+\Delta \varepsilon ]$ can be
written as $\mu _{0}=g\mu _{B}\left( B_{c}-B\right) $, which defines a
saturation field $B_{c}=W/\left( g\mu _{B}\right) $. Here the magnon
bandwidth $W$ is a function of spin coupling parameters $J,J^{\prime
},J^{\prime \prime }$ and $\mathcal{D}$. In terms of $J$, $W=2.898J$ and
therefore $B_{c}=8.51$T, assuming $g=2.2$.

Due to the symmetry property of two magnon branches $E_{\mathbf{Q}_{1}+%
\mathbf{q}}^{\mathcal{A}}=E_{\mathbf{Q}_{2}-\mathbf{q}}^{\mathcal{B}}$,
their densities of states (DOS) coinside, $\rho ^{\mathcal{A}}\left(
E\right) =\rho ^{\mathcal{B}}\left( E\right) $, where%
\begin{equation}
\rho ^{\mathcal{A},\mathcal{B}}\left( E\right) =\frac{1}{\mathcal{N}}\sum_{%
\mathbf{q}}\delta \left( E-E_{\mathbf{q}}^{\mathcal{A},\mathcal{B}}\right) .
\label{eq14}
\end{equation}%
The upper end of the calculated DOS is at $E_{u}=W$. For our purposes, the
most important is the low-energy part of $\rho ^{\mathcal{A},\mathcal{B}%
}\left( E\right) $ depicted in Fig. 2. A critical point in the magnon
spectrum is clearly seen at $E=E^{\ast }\simeq 50$mK: 3D character of the
spectrum at $E<E^{\ast }$ changes into 2D one for $E>E^{\ast }$.
\begin{figure}[tbh]
\epsfig{file=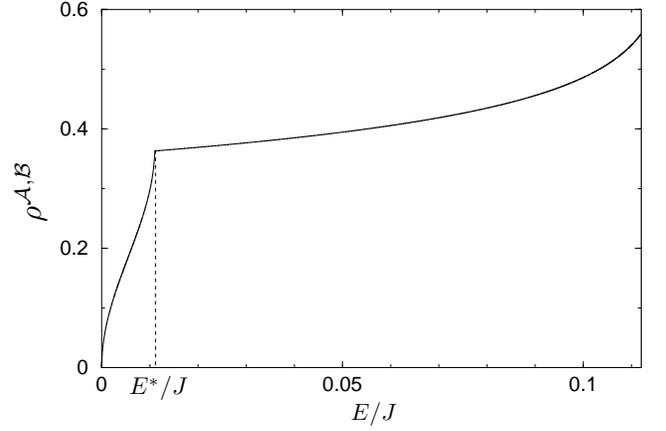,width=8.5cm}
\caption{Low-energy part of magnon density of states (in units 1/$J$) for $%
\mathcal{A}$ and $\mathcal{B}$ branches.}
\label{fig:2}
\end{figure}

\section{Effective magnon interaction}

When treating the magnon interactions, we take into account the dominant, $%
\mathcal{U}\rightarrow \infty $, hard-core repulsion and neglect effects of
a $\mathbf{q}$-dependent part of interaction. Then, $\mathcal{H}_{int}\simeq 
\mathcal{H}_{\mathcal{U}}^{\left( a\right) }+\mathcal{H}_{\mathcal{U}%
}^{\left( b\right) }$, where%
\begin{multline}
\mathcal{H}_{\mathcal{U}}^{\left( a\right) }+\mathcal{H}_{\mathcal{U}%
}^{\left( b\right) }= \\
\frac{\mathcal{U}}{\mathcal{N}}\sum_{\substack{ \mathbf{q}_{1},\mathbf{q}%
_{2},  \\ \mathbf{q}_{3},\mathbf{q}_{4}}}(a_{\mathbf{q}_{1}}^{+}a_{\mathbf{q}%
_{2}}^{+}a_{\mathbf{q}_{4}}a_{\mathbf{q}_{3}}+b_{\mathbf{q}_{1}}^{+}b_{%
\mathbf{q}_{2}}^{+}b_{\mathbf{q}_{4}}b_{\mathbf{q}_{3}})\Delta _{\mathbf{q}%
_{1}+\mathbf{q}_{2},\mathbf{q}_{3}+\mathbf{q}_{4}},  \label{eq15}
\end{multline}%
and $\Delta _{\mathbf{q}_{1}+\mathbf{q}_{2},\mathbf{q}_{3}+\mathbf{q}_{4}}$
implies the momentum conservation $\mathbf{q}_{1}+\mathbf{q}_{2}=\mathbf{q}%
_{3}+\mathbf{q}_{4}$. Below we refer to $\mathcal{H}_{\mathcal{U}}^{\left(
a\right) }$ and $\mathcal{H}_{\mathcal{U}}^{\left( b\right) }$ as the
hard-core magnon repulsion in $a$ and $b$ channels, respectively.

In high magnetic field, $B>B_{c}$, a fully spin polarized state is the
vacuum state for magnons. Upon decreasing field, in the very vicinity of $%
B_{c}$, i.e. $B\lesssim B_{c}$ and at low temperature the number of magnons
is very small, $n \sim \left(1-B/B_{c}\right)$. In this case an effective
magnon interaction can be found as a result of multiple magnon scattering by
summing up ladder diagrams.\cite{Abrikosov} When accomplishing such a
summation, we neglect interference between $a$ and $b$ channels. In this
approximation, the problem reduces to solving the similar Bethe-Salpether
equation for the renormalized scattering amplitudes $\Gamma^{\left( a\right)
}$ and $\Gamma^{\left( b\right) }$ in both channels. Therefore, below we
discuss with more detail the channel $a$.

By substituting the transformation $a_{\mathbf{q}}=\alpha _{\mathbf{q}}%
\mathcal{A}_{\mathbf{q}}-\beta _{\mathbf{q}}\mathcal{B}_{\mathbf{q}}$ into
interaction Hamiltonian $\mathcal{H}_{\mathcal{U}}^{\left( a\right) }$ in
Eq. (\ref{eq15}) one finds that $\mathcal{H}_{\mathcal{U}}^{\left( a\right)
} $ splits into sixteen scattering terms. In each process, the bare
interaction acquires now an extra factor given by a product of four $\alpha
_{\mathbf{q}}$ and $\beta _{\mathbf{q}}$ coefficients. For instance, a
scattering process $\left( \mathcal{A}_{\mathbf{q}_{3}},\mathcal{B}_{\mathbf{%
q}_{4}}\right) \rightarrow \left( \mathcal{A}_{\mathbf{q}_{1}},\mathcal{B}_{%
\mathbf{q}_{2}}\right) $ is described by the amplitude $2\alpha _{\mathbf{q}%
_{1}}\beta _{\mathbf{q}_{2}}\alpha _{\mathbf{q}_{3}}\beta _{\mathbf{q}_{4}}%
\mathcal{U}$. Taking into account multiple magnon scattering in the ladder
approximation leads to a replacement of $2\mathcal{U}$ by an effective
two-particle interaction $\Gamma ^{\left( a\right) }(\mathbf{K},\omega )$,
where $\bf{K}=\mathbf{q}_{1}+\mathbf{q}_{2}=\mathbf{q}_{3}+\mathbf{q}_{4}$
and $\omega =\omega _{1}+\omega _{2}=\omega _{3}+\omega _{4}$ are the total
momentum and energy of a scattering magnon pair; the extra factor remains
unchanged. The same holds for the other fifteen scattering processes. In the
magnetic field close to $B_{c}$ and at low $T$ only the particles near the
magnon band minima at $\mathbf{q}=\mathbf{Q}_{1,2}$ are present, which
allows us to send $\omega \rightarrow 0$ and consider $\Gamma ^{\left(
a\right) }(\mathbf{K},0)\equiv \Gamma _{\mathbf{K}}^{\left( a\right) }$. 
\begin{figure}[tbh]
\epsfig{file=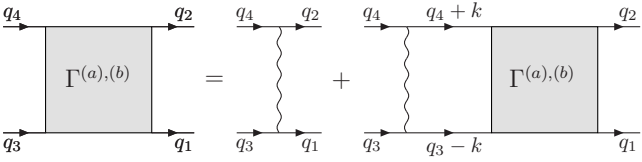,width=8.5cm}
\caption{Bethe-Salpeter equation for the scattering vertices $\Gamma
^{(a),(b)}$ }
\label{fig:3}
\end{figure}

The effective interaction $\Gamma _{\mathbf{K}}^{\left( a\right) }$ is
obtained in an explicit form by solving the Bethe-Salpeter equation
presented in Fig. 3 diagrammatically. In this presentation, $q_{i}=\left( 
\mathbf{q}_{i},\omega _{i}\right) $ and $k=(\bf{k},\omega ^{\prime })$ are
four-dimensional vectors. Any external incoming or outgoing line is
associated with either $\mathcal{A}$ or $\mathcal{B}$ magnon; two internal
lines labelled by $q_{4}+k$ and $q_{3}-k$ correspond to Greens's function $%
\mathcal{G}^{\left( a\right) }\left( q\pm k\right) =\alpha _{q\pm k}^{2}%
\mathcal{G}^{\left( \mathcal{A}\right) }\left( q\pm k\right) +\beta _{q\pm
k}^{2}\mathcal{G}^{\left( \mathcal{B}\right) }\left( q\pm k\right) $. Here $%
\mathcal{G}^{\left( \mathcal{A}\right) }\left( q\pm k\right) $ and $\mathcal{%
G}^{\left( \mathcal{B}\right) }\left( q\pm k\right) $ are the Green's
functions of decoupled $\mathcal{A}$ and $\mathcal{B}$ magnons. Since the
bare interaction $\mathcal{U}$ is $\mathbf{q}$- and $\omega $-independent,
the solution of the Bethe-Salpeter equation in the zero frequency limit is%
\begin{eqnarray}
\lbrack \Gamma _{\mathbf{q}_{1}+\mathbf{q}_{2}}^{\left( a\right) }]^{-1}
&=&\qquad \notag  \\
&&\frac{1}{\mathcal{N}}\sum_{\mathbf{k}}[\frac{\alpha _{\mathbf{q}_{1}+%
\mathbf{k}}^{2}\alpha _{\mathbf{q}_{2}-\mathbf{k}}^{2}}{E_{\mathbf{q}_{1}+%
\mathbf{k}}^{\mathcal{A}}+E_{\mathbf{q}_{2}-\mathbf{k}}^{\mathcal{A}}}+\frac{%
\alpha _{\mathbf{q}_{1}+\mathbf{k}}^{2}\beta _{\mathbf{q}_{2}-\mathbf{k}}^{2}%
}{E_{\mathbf{q}_{1}+\mathbf{k}}^{\mathcal{A}}+E_{\mathbf{q}_{2}-\mathbf{k}}^{%
\mathcal{B}}}  \notag \\
&&+\frac{\beta _{\mathbf{q}_{1}+\mathbf{k}}^{2}\alpha _{\mathbf{q}_{2}-%
\mathbf{k}}^{2}}{E_{\mathbf{q}_{1}+\mathbf{k}}^{\mathcal{B}}+E_{\mathbf{q}%
_{2}-\mathbf{k}}^{\mathcal{A}}}+\frac{\beta _{\mathbf{q}_{1}+\mathbf{k}%
}^{2}\beta _{\mathbf{q}_{2}-\mathbf{k}}^{2}}{E_{\mathbf{q}_{1}+\mathbf{k}}^{%
\mathcal{B}}+E_{\mathbf{q}_{2}-\mathbf{k}}^{\mathcal{B}}}].  \label{eq16}
\end{eqnarray}%
The similar derivation applies also to an effective magnon interaction $%
\Gamma _{\mathbf{q}_{1}+\mathbf{q}_{2}}^{\left( b\right) }$ in the channel $b
$, which can be obtained merely from $\Gamma _{\mathbf{q}_{1}+\mathbf{q}%
_{2}}^{\left( a\right) }$ by interchanging $\mathcal{A}\leftrightarrow 
\mathcal{B}$.

By collecting back all the terms in both $a$ and $b$ channels, one arrives
at an effective interaction Hamiltonians of magnons. It becomes clear now
that the effective Hamiltonian takes the original form (\ref{eq15}) where
the bare parameter $2\mathcal{U}$ has to be replaced by $\Gamma _{\mathbf{q}%
_{1}+\mathbf{q}_{2}}^{\left( a\right) }$ and $\Gamma _{\mathbf{q}_{1}+%
\mathbf{q}_{2}}^{\left( b\right) }$ in channel $a$ and $\ b,$ respectively.

\section{Mean-field theory}

The main goal of the present study is to calculate a shape of the phase
boundary, i.e. a field dependence of the critical temperature $T_{c}\left(
B\right) $ for $B\lesssim B_{c}$. When approaching $B_{c}$ from below, $%
T_{c}\rightarrow 0$. In the low-$T$ region near $B_{c}$, only low-energy
magnon states at the band minima wave-vectors $\mathbf{q}=\mathbf{Q}_{1},%
\mathbf{Q}_{2}$ are occupied and contribute to the phase transition. In this
section, we consider first how the magnon spectrum near the band minima is
renormalized due to the effective magnon interaction. The study is based on
the Hartree-Fock (HF) approximation, since near the quantum critical point, $%
B=B_{c}$ and $T=0$, the HF theory is suggested to describe correctly the
phase boundary.\cite{Nohadani04,Kawashima04}

We assume that the temperature approaches $T_{c}$ from above. Then in the
disordered phase, the mean-field interaction Hamiltonian in the channel $a$
can be written as%
\begin{equation}
\mathcal{H}_{int,MF}^{\left( a\right) }\simeq \frac{2}{\mathcal{N}}\sum_{%
\mathbf{q}_{1},\mathbf{q}_{2}}\Gamma _{\mathbf{q}_{1}+\mathbf{q}%
_{2}}^{\left( a\right) }\left\langle a_{\mathbf{q}_{1}}^{+}a_{\mathbf{q}%
_{1}}\right\rangle a_{\mathbf{q}_{2}}^{+}a_{\mathbf{q}_{2}},  \label{eq17}
\end{equation}%
which is approximated further as follows%
\begin{multline}
\Gamma _{\mathbf{q}_{1}+\mathbf{q}_{2}}^{\left( a\right) }\left\langle a_{%
\mathbf{q}_{1}}^{+}a_{\mathbf{q}_{1}}\right\rangle \simeq \Gamma _{\mathbf{Q}%
_{1}+\mathbf{q}_{2}}^{\left( a\right) }\alpha _{\mathbf{Q}%
_{1}}^{2}\left\langle \mathcal{A}_{\mathbf{q}_{1}}^{+}\mathcal{A}_{\mathbf{q}%
_{1}}\right\rangle   \\
+\Gamma _{\mathbf{Q}_{2}+\mathbf{q}_{2}}^{\left( a\right) }\beta _{\mathbf{Q}%
_{2}}^{2}\left\langle \mathcal{B}_{\mathbf{q}_{1}}^{+}\mathcal{B}_{\mathbf{q}%
_{1}}\right\rangle \\
-\alpha _{\mathbf{q}_{1}}\beta _{\mathbf{q}_{1}}\Gamma _{\mathbf{q}_{1}+%
\mathbf{q}_{2}}^{\left( a\right) }\left\langle \mathcal{A}_{\mathbf{q}%
_{1}}^{+}\mathcal{B}_{\mathbf{q}_{1}}+\mathcal{B}_{\mathbf{q}_{1}}^{+}%
\mathcal{A}_{\mathbf{q}_{1}}\right\rangle .  \label{eq18}
\end{multline}%
In this expression, we made replacements $\Gamma _{\mathbf{q}_{1}+\mathbf{q}%
_{2}}^{\left( a\right) }\alpha _{\mathbf{q}_{1}}^{2}\rightarrow \Gamma _{%
\mathbf{Q}_{1}+\mathbf{q}_{2}}^{\left( a\right) }\alpha _{\mathbf{Q}_{1}}^{2}
$ and $\Gamma _{\mathbf{q}_{1}+\mathbf{q}_{2}}^{\left( a\right) }\beta _{%
\mathbf{q}_{1}}^{2}\rightarrow \Gamma _{\mathbf{Q}_{2}+\mathbf{q}%
_{2}}^{\left( a\right) }\beta _{\mathbf{Q}_{2}}^{2}$, since the low-energy
magnon densities $\left\langle n_{\mathbf{q}}^{\mathcal{A}}\right\rangle
=\left\langle \mathcal{A}_{\mathbf{q}}^{+}\mathcal{A}_{\mathbf{q}%
}\right\rangle $ and $\left\langle n_{\mathbf{q}}^{\mathcal{B}}\right\rangle
=\left\langle \mathcal{B}_{\mathbf{q}}^{+}\mathcal{B}_{\mathbf{q}%
}\right\rangle \ $are located in the $\mathbf{q}$-space near $\mathbf{Q}_{1}$
and $\mathbf{Q}_{2}$ for $\mathcal{A}$ and $\mathcal{B}$ branch,
respectively. Because of a symmetry of the magnon interaction, the band
minima degeneracy is preserved and one has $\frac{1}{\mathcal{N}}\sum_{%
\mathbf{q}}\left\langle n_{\mathbf{q}}^{\mathcal{A}}\right\rangle =\frac{1}{%
\mathcal{N}}\sum_{\mathbf{q}}\left\langle n_{\mathbf{q}}^{\mathcal{B}%
}\right\rangle =n$. A possible shift due to magnon interaction of the band
minima location in the $\mathbf{q}$-space at $\vec{Q}_{1,2}$ is assumed to
be small and neglected below. In the zero-order approximation , the
quantities $\left\langle \mathcal{A}_{\mathbf{q}_{1}}^{+}\mathcal{B}_{%
\mathbf{q}_{1}}+\mathcal{B}_{\mathbf{q}_{1}}^{+}\mathcal{A}_{\mathbf{q}%
_{1}}\right\rangle $ vanish, and we omit them in Eq. (\ref{eq18}) as well. A
validity of the latter approximation will be verified later on.

Similar arguments as above allow us to approximate the operator part in Eq. (%
\ref{eq17}) as follows%
\begin{eqnarray}
a_{\mathbf{q}}^{+}a_{\mathbf{q}} &\simeq &\alpha _{Q_{1}}^{2}\mathcal{A}_{%
\mathbf{q}}^{+}\mathcal{A}_{\mathbf{q}}+\beta _{Q_{2}}^{2}\mathcal{B}_{%
\mathbf{q}}^{+}\mathcal{B}_{\mathbf{q}}  \notag \\
&&-\alpha _{\mathbf{q}}\beta _{\mathbf{q}}\left( \mathcal{A}_{\mathbf{q}}^{+}%
\mathcal{B}_{\mathbf{q}}+\mathcal{B}_{\mathbf{q}}^{+}\mathcal{A}_{\mathbf{q}%
}\right) ,  \label{eq19}
\end{eqnarray}%
and to treat the effective magnon interaction in the channel $b$ in the same
manner as in the channel $a$. Finally, by noting that $\alpha _{\mathbf{Q}%
_{1}}^{2}=\alpha _{\mathbf{Q}_{2}}^{2}=\alpha ^{2}$, $\beta _{\mathbf{Q}%
_{1}}^{2}=\beta _{\mathbf{Q}_{2}}^{2}=\beta ^{2}$ and $\alpha ^{2}\gg \beta
^{2}$, we retain only the leading terms in the mean-field interaction
Hamiltonian, which leads to the following result%
\begin{multline}
\mathcal{H}_{int,MF}^{\left( a\right) }+\mathcal{H}_{int,MF}^{\left(
b\right) }=   \\
2\Gamma n\sum_{\mathbf{q}}\left( \mathcal{A}_{\mathbf{q}}^{+}\mathcal{A}_{%
\mathbf{q}}+\mathcal{B}_{\mathbf{q}}^{+}\mathcal{B}_{\mathbf{q}}\right)  
 \\
-2n\sum_{\mathbf{q}}\Gamma _{\mathbf{q}}^{^{\prime }}\left( \mathcal{A}_{%
\mathbf{q}}^{+}\mathcal{B}_{\mathbf{q}}+\mathcal{B}_{\mathbf{q}}^{+}\mathcal{%
A}_{\mathbf{q}}\right) ,  \label{eq20}
\end{multline}%
where%
\begin{equation}
\begin{array}{c}
\Gamma =\alpha ^{4}\Gamma _{2\mathbf{Q}_{1}}^{\left( a\right) }=\alpha
^{4}\Gamma _{2\mathbf{Q}_{2}}^{\left( b\right) } \\ 
\\ 
\Gamma _{\mathbf{q}}^{^{\prime }}=\alpha _{\mathbf{q}}\beta _{\mathbf{q}%
}[\Gamma _{\mathbf{Q}_{1}+\mathbf{q}}^{\left( a\right) }-\Gamma _{\mathbf{Q}%
_{2}+\mathbf{q}}^{\left( b\right) }]%
\end{array}
\label{eq21}
\end{equation}%
According to Eq. (\ref{eq20}), the bare chemical potential of magnons is
renormalized%
\begin{equation}
\mu _{0}\rightarrow \mu _{eff}=\mu _{0}-2\Gamma n.  \label{eq22}
\end{equation}%
With the use of the definition (\ref{eq12}) for $\alpha _{\mathbf{q}}$ and (%
\ref{eq16}) for $\Gamma ^{\left( a\right) }$, we obtain numerically the
estimate $\Gamma \simeq 0.85J$. The second term in Eq. (20) generates magnon
hybridization. This term shifts the bottom of the magnon bond slightly down
and leads to a weak mass enhancment of low-energy magnons. Both effects are
proportional to $n^{2}$ and we omit them since $n\ll 1$ near $T_{c}$ for $%
(B_{c}-B)\ll B_{c}$. In this region, the low-energy magnon $\mathcal{A}$ and 
$\mathcal{B}$ branches remain decoupled, which justifies an approximation
made before.

\section{Critical temperature $T_{c}\left( B\right) $ near $B_{c}$}

In the previous sections we obtained in the dilute magnon limit $n\ll1$, and
within the HF approximation that the leading, linear in $n$, effect of
magnon interaction is a renormalization of the chemical potential, Eq. (\ref%
{eq22}). In this section we calculate the critical temperature $T_{c}\left(
B\right) $ of a magnon BEC transition as a function of \thinspace$B$
assuming that for given $B\lesssim B_{c}$ the temperature approaches the
phase boundary from the normal phase.

\begin{figure}[tbh]
\epsfig{file=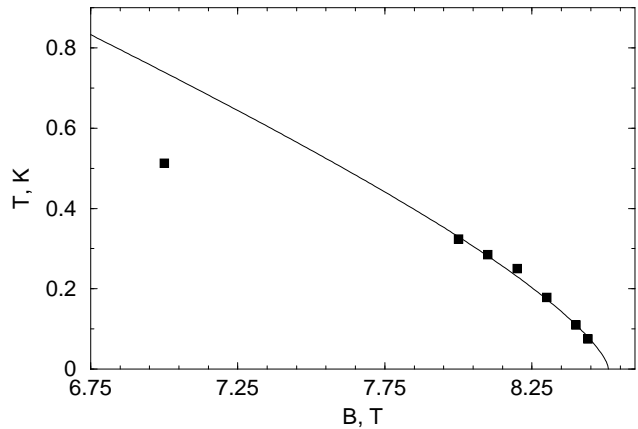,width=8.5cm} .
\caption{Calculated phase boundary $T_{c}\left( B\right)$ near $B_c$ (solid
line). Experimental data (squares) are from specific heat measurements
reported in Ref. \onlinecite{Radu05}}
\label{fig:5}
\end{figure}

As follows from Eq.(\ref{eq22}), a comparatively large magnon density $n$ at
high temperature makes the chemical potential to be negative, $\mu _{eff}<0$%
, and all the magnons are noncondensed. Under decreasing $T$, $\mu _{eff}$
tends to zero and a BEC transition is expected to occur when $\mu _{eff}=0$
at $T_{c}\left( B\right) $; for $T\leqslant T_{c}$, a condensate density $%
n_{0}$ starts to develop continuously at the magnon band minima. From Eq. (%
\ref{eq22}), the condition for the phase transition is written as\cite%
{Nikuni00}%
\begin{equation}
g\mu _{B}\left( B_{c}-B\right) =2\Gamma n\left( T_{c}\right) .  \label{eq23}
\end{equation}%
Here $n\left( T\right) =\mathcal{N}^{-1}\sum_{\mathbf{q}}f_{B}\left( E_{%
\mathbf{q}}\right) $ with $f_{B}\left( E_{\mathbf{q}}\right) $ being the
Bose distribution function taken at $\mu _{eff}=0$ and $E_{\mathbf{q}}=E_{%
\mathbf{q}}^{\mathcal{A}}$ or $E_{\mathbf{q}}=E_{\mathbf{q}}^{\mathcal{B}}.$
This means that for $T<T_{c}$ the magnon condensate develops simultaneously
at $\mathbf{q}=\mathbf{Q}_{1,2}$. Since at $\mu _{eff}=0$ the distribution
function $f_{B}\left( E\right) $ diverges as $T/E$ for $E\rightarrow 0$, a
magnon spectrum near the band minima mainly contribute to $n\left(
T_{c}\right) $. If $T_{c}$ is sufficiently low, the very low-energy magnons
having a 3D quadratic dispersion dominate and drive the BEC transition. In
this asymptotic regime the power-law is expected\ \cite%
{Giamarchi99,Nohadani04,Kawashima04} $T_{c}\left( B\right) \sim \left(
B_{c}-B\right) ^{1/\phi }$ with the universal exponent $\phi _{BEC}=3/2$. To
check this suggestion and to compare the calculated phase boundary with that
determined experimentally in Cs$_{2}$CuCl$_{4}$, we solved the equation (\ref%
{eq23}) with the use of the magnon density of states, Fig. 2, realistic for
Cs$_{2}$CuCl$_{4}$. The result of calculations, with $B_{c}=8.51$T and $%
\Gamma =0.85J$, is shown in Fig. 4 and compared with experimental data\cite%
{Radu05} for $B\lesssim B_{c}$. In the very vicinity to $B_{c}$, the
theoretical curve is well reproduced by the power-law with the calculated $%
\phi _{th}\simeq 1.5$, while a fit of experimental data yields\cite{Radu05} $%
\phi _{exp}=1.52(10)$; both values of the critical exponent are close to the
universal $\phi _{BEC}=3/2$.\newline

\section{Conclusions}

Based on a realistic quantum spin model in the hard-core boson
representation, we developed a mean-field theory of a magnon BEC transition
in Cs$_{2}$CuCl$_{4}$ near the saturation field $B_{c}$. The calculated
critical temperature $T_{c}\left( B\right) $ describes surprisingly well the
experimentally measured phase boundary in a narrow field region $%
B_{c}-\Delta B<B<B_{c}$, where $\Delta B/B_{c}\simeq 0.06$. In this region,
where $0\leqslant T_{c}<300$mK, both the measured and calculated critical
exponents $\phi $ are very close to the universal value $\phi _{BEC}=3/2$
characteristic for 3D quadratic magnon dispersion. It means that in Cs$_{2}$%
CuCl$_{4}$ the magnon BEC near $B_{c}$ is driven by 3D magnon states with
energies below $E^{\ast }\approx 50$mK. Actually, by sending formally $%
E^{\ast }\rightarrow 0$ (see Fig. 2) in our calculations we obtained $%
T_{c}=0 $ at any $B.$ At lower fields $B<B_{c}-\Delta B\simeq 8$T, a
mean-field description of the experimentally determined phase boudary $%
T_{c}\left( B\right) $ fails, as was expected. Both thermal fluctuations and
a temperature renormalization of the magnon spectrum have to be taking into
account to improve a theoretical description of the phase boundary at lower
magnetic fields.

\section{Acknowledgments}

We are grateful to D.Ihle, P.Thalmeier, M. Sigrist, and M.Vojta for many
stimulating discussions. D.K. thanks G.V.Pai and S.Sinha for helpful
discussions. V.Y. acknowledges financial support by Deutsche
Forschungsgemeinschaft.

\end{document}